\documentclass[floats,floatfix,showpacs,amssymb,prd,twocolumn,superscriptaddress,nofootinbib,nolongbibliography,reprint]{revtex4-2}

\usepackage{amssymb,amsmath,verbatim,mathtools,needspace,enumitem,etoolbox,graphicx,physics,microtype,afterpage,bigints,placeins}
\usepackage{gensymb}

\usepackage[dvipsnames, usenames]{xcolor}

\definecolor{linkcolor}{rgb}{0.0,0.3,0.5}
\usepackage[unicode, colorlinks=true, linkcolor=linkcolor, citecolor=linkcolor, filecolor=linkcolor,urlcolor=linkcolor, pdfusetitle]{hyperref}
\usepackage[all]{hypcap}
\usepackage[T1]{fontenc}
\usepackage[utf8]{inputenc}
\usepackage{tabularx}
\usepackage{lmodern}

\newcommand{\jhu}{\affiliation{Department of Physics and Astronomy, Johns Hopkins University, 3400 N. Charles Street, Baltimore, Maryland 21218, USA}}
\newcommand{\bham}{\affiliation{School of Physics and Astronomy \& Institute for Gravitational Wave Astronomy, \\ University of Birmingham, Birmingham, B15 2TT, UK}}

\usepackage{lipsum}

\def\nn{\nonumber}

\renewcommand{\vec}[1]{\boldsymbol{#1}}

\newcommand{\ben}{\begin{enumerate}}
\newcommand{\een}{\end{enumerate}}

\def\be{\begin{equation}}
\def\ee{\end{equation}}
\def\bea{\begin{eqnarray}}
\def\eea{\end{eqnarray}}
\def\nn{\nonumber}
\newcommand{\beq}{\begin{eqnarray}}
\newcommand{\eeq}{\end{eqnarray}} 
\newcommand{\ba}{\begin{align}}
\newcommand{\ea}{\end{align}}

\definecolor{royalblue4}{HTML}{27408B}

\def\nn{\nonumber}
\def\be{\begin{equation}}
\def\ee{\end{equation}}
\def\beq{\begin{eqnarray}}
\def\eeq{\end{eqnarray}}
\def\f{\frac}

\newcommand{\chiparent}{\xi}
\newcommand{\chionep}{\xi_1}
\newcommand{\chitwop}{\xi_2}
\newcommand{\chirem}{\chi_\mathrm{r}}

\newcommand{\qp}{Q}

\newcommand{\Msun}{\,{\rm M}_{\odot}}

\newcommand{\chimax}{\chi_{\rm max}}
\newcommand{\chieff}{\chi_{\rm eff}}
\newcommand{\chip}{\chi_{\rm p}}

\newcommand{\ssim}{\mathchar"5218\relax\,}

\newcommand{\chit}{\tilde{\xi}}

\newcolumntype{L}{>{$}l<{$}}
\newcolumntype{R}{>{$}r<{$}}
\newcolumntype{C}{>{$}c<{$}}

\def\nn{\nonumber}
\def\be{\begin{equation}}
\def\ee{\end{equation}}
\def\beq{\begin{eqnarray}}
\def\eeq{\end{eqnarray}}
\def\f{\frac}

\begin{document}

\title{Looking for the parents of LIGO's black holes}

\author{Vishal Baibhav}
\email{vbaibha1@jhu.edu}
\jhu
\author{Emanuele Berti}
\jhu
\author{Davide Gerosa}
\bham
\author{Matthew Mould}
\bham
\author{Kaze W. K. Wong}
\jhu
\date{\today}

\begin{abstract}
Solutions to the two-body problem in general relativity allow us to predict the mass, spin, and recoil velocity of a black-hole merger remnant given the masses and spins of its binary progenitors. We address the inverse problem: given a binary black-hole merger, can we use the parameters measured by gravitational-wave interferometers to determine whether the binary components are of hierarchical origin, i.e. whether they are themselves remnants of previous mergers? If so, can we determine at least some of the properties of their parents? This inverse problem is in general overdetermined. We show that hierarchical mergers occupy a characteristic region in the plane composed of the effective-spin parameters $\chi_{\rm eff}$ and $\chi_{\rm p}$, and therefore a measurement of these parameters can add weight to the hierarchical-merger interpretation of some gravitational-wave events, including GW190521. If one of the binary components has hierarchical origin and its spin magnitude is well measured, we derive exclusion regions on the properties of its parents: for example we infer that the parents of GW190412 (if hierarchical) must have had
unequal masses
and low spins. Our formalism is quite general, and it can be used to infer constraints on the astrophysical environment producing hierarchical mergers.
\end{abstract}

\maketitle

\begin{figure*}
\includegraphics[trim=15.5cm 2cm 13cm 9cm,width=0.65\textwidth,clip]{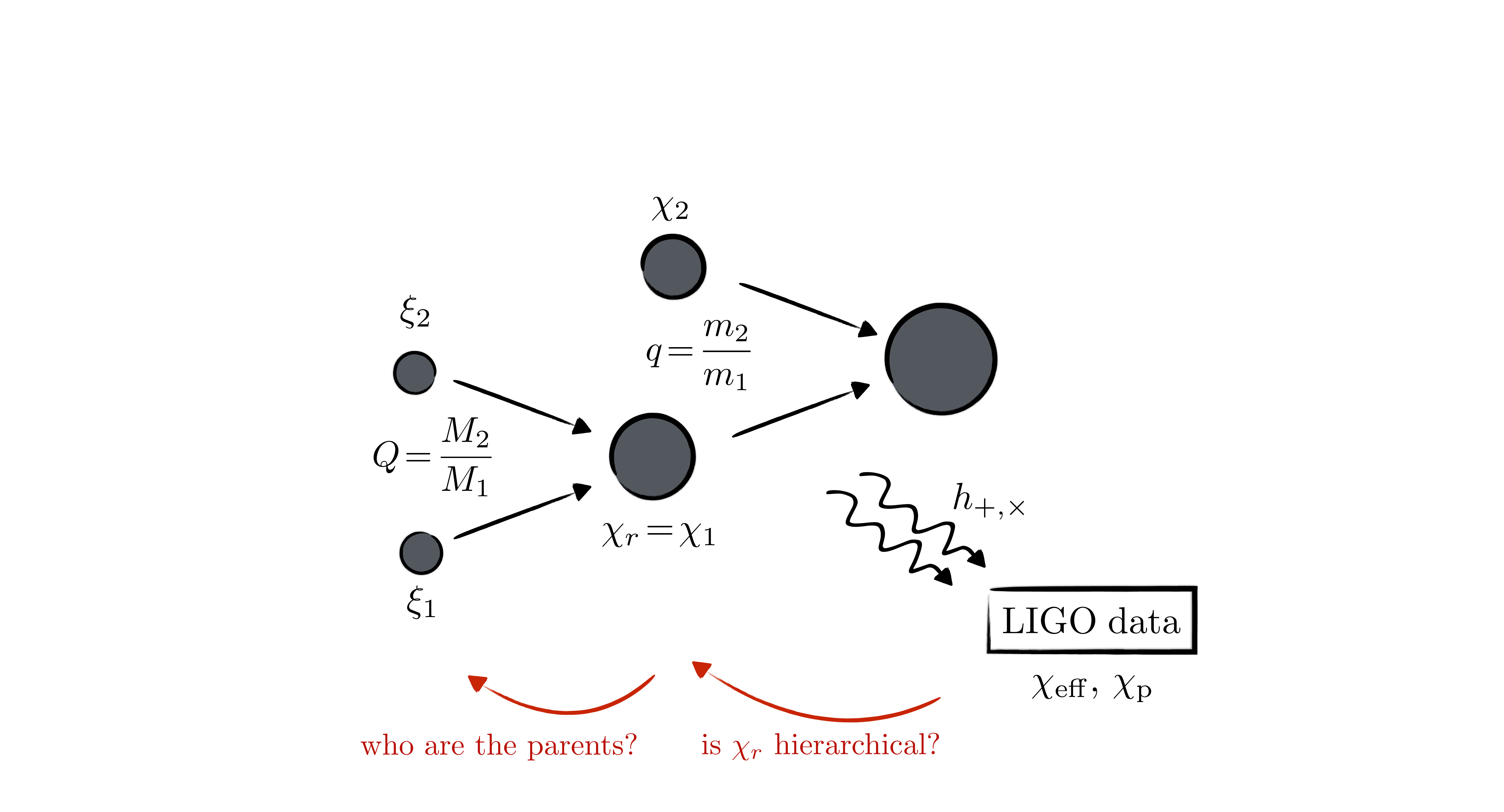}
\caption{Schematic representation of the inverse problem addressed in this paper. In Sec.~\ref{chieffchipplane} we ask: if we detect GWs from a binary BH merger with mass ratio $q$, effective spin $\chieff$ and precessional spin $\chip$, can we establish whether one of the binary components (in this sketch, the one with remnant spin $\chirem=\chi_1$) originated from a previous merger?  In Sec.~\ref{parents} we instead ask: if indeed one of the binary components comes from a previous merger and we measure $\chirem$, can we determine the mass ratio $\qp$, spin magnitudes $\xi_{1,2}$ and recoil velocity of the parent binary? The merger remnant does not necessarily have to be the primary component of the observed binary, as shown in this sketch (i.e., the arguments in Sec.~\ref{parents} also apply to the case where $\chirem=\chi_2$). %
}
\label{cartoonparents}
\end{figure*}

\section{Introduction}

Gravitational-wave (GW) astronomy is rapidly entering a data-driven regime where astrophysical modeling uncertainties are becoming the key limiting factor. 
The most popular astrophysical formation channels for the merging black-hole (BH) binaries detected by LIGO and Virgo include isolated binary evolution in the galactic field and dynamical assembly in either dense stellar clusters or accretion disks~\cite{Mandel:2018hfr,Mapelli:2018uds}. %
Astrophysical predictions depend on several poorly understood phenomena (nuclear-reaction rates, supernova kicks, common-envelope efficiency, metallicity, formation and evolution of stellar clusters), which make the various channels largely or partially degenerate. %
While there is some evidence that multiple formation channels provide comparable contributions to the overall rate of merger events detectable in GWs~\cite{Callister:2020vyz,Zevin:2020gbd}, pinpointing the origin of specific systems observed by LIGO and Virgo remains challenging. 

The difficulty of the task is greatly alleviated by the so-called gaps in the BH binary parameter space: regions that can be populated only by a subset of the proposed formation channels. The mass spectrum of merging BHs is predicted to present two such gaps. The typical timescales involved in the explosion mechanism may prevent the formation of compact objects between $\sim 3 M_\odot$ and $\sim 5 M_\odot$~\cite{Fryer:2011cx}, while unstable pair production during the advanced evolutionary stages of massive stars can impede collapse to BHs with masses $\gtrsim 50 M_\odot$ and $\lesssim 120 M_\odot$~\cite{Farmer:2019jed,Woosley:2021xba}. The existence of both gaps is partially supported by current observations, which show that the BH binary merger rate is indeed suppressed in those regions~\cite{Abbott:2020gyp}. However, observations also tell us that the gaps are somehow polluted, with GW190814~\cite{Abbott:2020khf} and GW190521~\cite{Abbott:2020tfl} presenting component masses that sit squarely in the lower and upper mass gap, respectively.%

Spin orientations are often invoked as a promising tool to distinguish formation channels, with binaries born in the field (assembled dynamically) being more likely to enter the LIGO band with small (large) spin-orbit misalignments~\cite{Gerosa:2013laa,Rodriguez:2016vmx,Stevenson:2017dlk,Gerosa:2018wbw}.  However, the discriminating power of the spin directions trivially fades if the spin magnitudes turn out to be vanishingly small, as predicted by some~\cite{Fuller:2019sxi,Belczynski:2017gds} (but not all~\cite{Zaldarriaga:2017qkw,Steinle:2020xej}) models of stellar evolution. In analogy with the mass gaps highlighted above, we previously referred to the putative absence of high-spinning BHs as the ``spin gap''~\cite{Baibhav:2020xdf}.  While the bulk of the observed population appears to be compatible with small but nonzero spins~\cite{Abbott:2020gyp}, both GW190521~\cite{Abbott:2020tfl}  and GW190412~\cite{LIGOScientific:2020stg} present strong evidence for spin dynamics.%

Hierarchical BH mergers are a natural strategy to populate both the upper mass gap and the spin gap~\cite{Gerosa:2017kvu,Fishbach:2017dwv} (see \cite{Gerosa:2021mno} for a review). By avoiding stellar collapse altogether, assembling objects from older generations of BH mergers bypasses the constraint imposed by both pair production instabilities and core-envelope interactions. The masses of merger remnants are roughly equal to the sum of the masses of the merging BHs, while their spins follow a  characteristic distribution that is highly peaked at  $\sim 0.7$. The key prediction is that of a positive correlation between masses and spins, with a depleted region in the high-mass/low-spin corner of the parameter space~\cite{Gerosa:2021hsc,Tagawa:2021ofj}. 
Hierarchical mergers could explain the mass/spin gap properties of both GW190412~\cite{Gerosa:2020bjb,Rodriguez:2020viw} and GW190521~\cite{Abbott:2020mjq}. Some population analyses are tentatively reporting that a subpopulation of hierarchical mergers might be present in the data~\cite{Kimball:2020qyd,Tiwari:2020otp,Baxter:2021swn}, although current evidence is inconclusive~\cite{Abbott:2020gyp}. 
In order to host hierarchical mergers, astrophysical environments need to possess a sufficiently large escape speed to retain merger remnants following relativistic recoils~\cite{Gerosa:2019zmo}. Nuclear star clusters, accretion disks surrounding active galactic nuclei (AGN), and possibly globular clusters are among the most plausible hosts~\cite{OLeary:2016ayz,Rodriguez:2019huv,Yang:2019cbr,Baibhav:2020xdf,Tagawa:2019osr,Mapelli:2021syv,Fragione:2020nib}.

This line of reasoning leads to two deeply connected questions, which are illustrated schematically in Fig.~\ref{cartoonparents}. First, do the measured parameters of a GW event allow us to tell whether it originated hierarchically from previous mergers? If so, what are the properties of the parent BHs?
While the main goal of LIGO/Virgo data analysis is to infer the properties of the merging BHs from the observed GWs, here we take one step backward in the BH family tree and try to constrain the properties of the parent BHs that generated the observed binary. This problem is in general overdetermined, but we show that it is possible to infer many properties of the parent binary.

Current spin inference is largely limited to estimates of some effective combinations of the two spins. In Sec.~\ref{chieffchipplane}, we show that hierarchical mergers occupy a distinct region in the plane composed of the two commonly used parameters $\chi_{\rm eff}$ and $\chi_{\rm p}$, which can thus be used to infer whether one or both of the merging BHs originated from a previous merger. We apply this argument to GW190521 and show that, together with its mass-gap properties, its spin values add considerable weight to a hierarchical-merger interpretation. Next, in Sec.~\ref{parents} we show that events with a well-constrained component spin magnitude allow for a unique reconstruction of the previous generation of BHs.
Our argument relies on inverting a numerical-relativity fit for the remnant spin. We apply this inversion to GW190412 and find that its parent binary must have had a %
moderate mass ratio and moderately low spins. 
Finally, in Sec.~\ref{concl} we present conclusions and directions for future work. We use geometrical units $G=c=1$ and GW posterior samples from  Ref.~\cite{Abbott:2020niy}.

\section{Hierarchical black-hole mergers in  the $(\boldsymbol\chi_{\bf eff},\,\boldsymbol\chi_{\bf p})$ plane}
\label{chieffchipplane}

A GW signal depends in principle on all 6 degrees of freedom corresponding to the two spin vectors $ \boldsymbol{\chi}_1$ and $ \boldsymbol{\chi}_2$, but most of the discriminating power is contained in a limited number of effective parameters. In this section we illustrate how, even with this limited set of information, one can draw powerful constraints on the likelihood that a given GW event originated from previous mergers.

\subsection{Effective spins}

The two spin components parallel to the binary's orbital angular momentum are often combined into the mass-weighted expression~\cite{Racine:2008qv}
\begin{align}\label{eq:chieff}
\chieff \equiv \frac{\chi_1\cos\theta_1 + q\chi_2\cos\theta_2}{1+q} \, ,
\end{align}
where $q=m_2/m_1\leq 1$ is the mass ratio, $\chi_i=|\vec{S}_i|/m_i$ %
 are the dimensionless Kerr parameters of the individual BHs, and $\theta_i = \arccos\small(\hat{\vec{S_i}}\cdot \hat{\vec{L}}\small)$ are the spin-orbit angles. %
The effective spin $\chieff$
is a constant of motion at second post-Newtonian (2PN) order~\cite{Racine:2008qv,Gerosa:2015tea}.%
 
The precession of the orbital plane is encoded in the variation of the direction of the orbital angular momentum $|{d\hat{\vec{L}}}/{dt}|$. It is usually described in terms of a spin parameter $\chi_{\rm p}$ first defined in a heuristic fashion as~\cite{Schmidt:2014iyl}
\begin{align}
\chi_{\rm p}^{\rm (heu)}= \max\left(\chi_{1\perp},  \chi_{2\perp}\right)\,,
\label{eq:chiph}
\end{align}
where 
\begin{align}
\chi_{1\perp} = \chi_1 \sin\theta_1\,,\quad
\chi_{2\perp} = q\frac{4q + 3}{4 + 3q}  \chi_2 \sin\theta_2\,,
\label{chipterms}
\end{align}
This parameter is routinely used in GW data analysis~\cite{Abbott:2020gyp}.

The quantity $\chi_{\rm p}^{\rm (heu)}\in [0,1]$ retains some, but not all, precession-timescale variations. This inconsistency was rectified
in Ref.~\cite{Gerosa:2020aiw}
using a complete precession average of $|{d\hat{\vec{L}}}/{dt}|$ at 2PN. Their augmented definition $\chi_{\rm p}^{\rm (av)}\in [0,2]$  has the desirable properties of (i) being conserved on the short spin-precession timescale of the problem, (ii) consistently including two-spin effects, (iii) agreeing with the heuristic expression in the single-spin limit. In the large-separation limit $r\to \infty$ (corresponding to GW emission frequencies $f_{\rm GW}\to 0$), the quantity $\chi_{\rm p}^{\rm (av)}$ can be written in closed form as \cite{Gerosa:2020aiw}
\begin{align}
\lim_{r\to\infty} \chip^{\rm (av)} &= 
\frac{|\chi_{1\perp} - \chi_{2\perp}|}{\pi}\,{\rm E} \!\left[-\frac{4\chi_{1\perp} \chi_{2\perp}}{(\chi_{1\perp} - \chi_{1\perp})^2}\right]
\notag\\&+ \frac{\chi_{1\perp} + \chi_{2\perp}}{\pi}\, {\rm E}\! \left[\frac{4\chi_{1\perp} \chi_{2\perp}}{(\chi_{1\perp} + \chi_{2\perp})^2}\right]
\label{ellipE}
\, ,
\end{align}
where ${\rm E}(m)$ is the complete elliptic integral of {the} second kind. In practice, this expression is accurate down to $\lesssim0.1\% \ (1\%)$ for $r\gtrsim10^6M \ (10^4M)$ (where $M=m_1+m_2$ is the total mass of the binary),
and thus well describes BH binaries at the large separations where they form. For configurations in the sensitivity window of LIGO/Virgo ($f_{\rm GW}\gtrsim 10$ Hz), one instead needs to use $ \chip^{\rm (av)} $ as given in Eq.~(16) of Ref.~\cite{Gerosa:2020aiw}.

\begin{figure*}[t]
  \includegraphics[width=\textwidth]{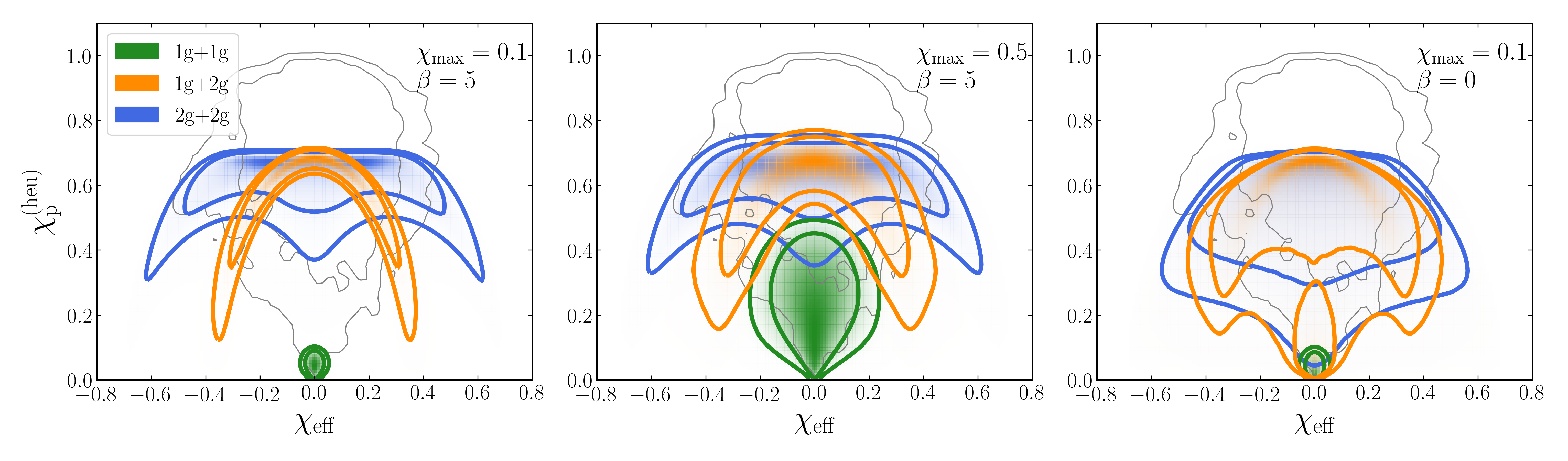}\\
  \includegraphics[width=\textwidth]{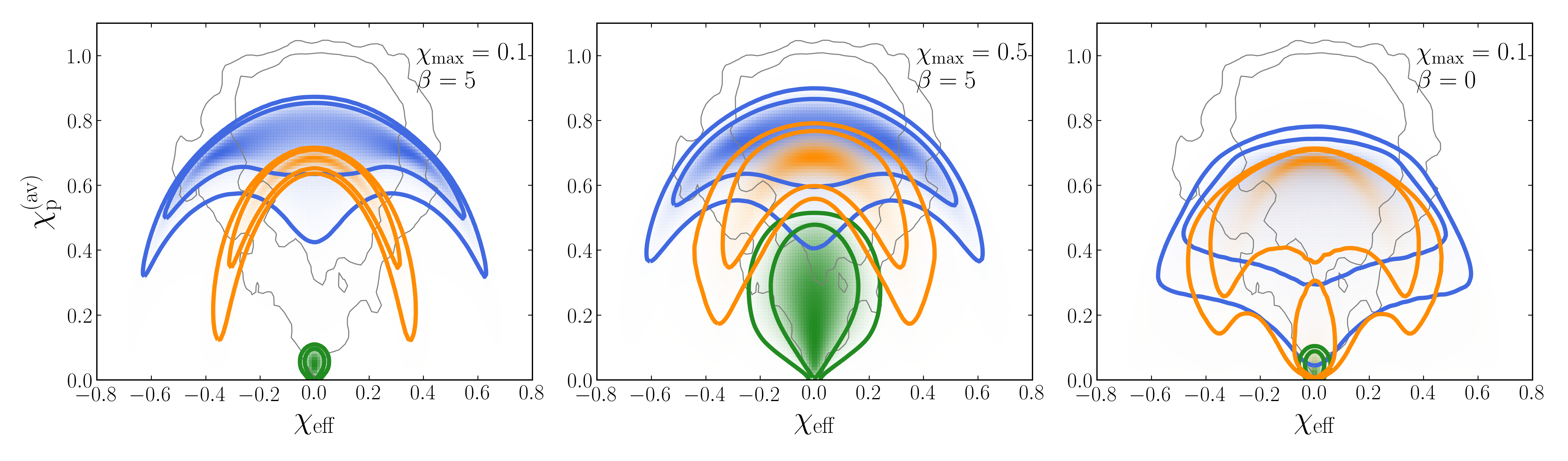}
  \caption{Distribution of different BH generations in  the $(\chieff,\,\chip)$ plane. Green, orange, and blue regions indicate 1g+1g, 1g+2g, and 2g+2g mergers, respectively, with contours marking the 68\% and 90\% levels.
  We consider a simple population model where $\chimax$ indicates the largest BH spins at formation, and $\beta$ indicates the binary pairing properties. In particular, we consider $(\chimax,\beta)=(0.1,5)$ (small spins and selective pairing, left panels), $(\chimax,\beta)=(0.5,5)$ (large spins and selective pairing, middle panels), and $(\chimax,\beta)=(0.1,0)$ (small spins and random pairing, right panels). We contrast the aligned effective spin $\chieff$ with the heuristic value of the precession parameter $\chip^{(\rm heu)}$ (bottom panels) and the asymptotic limit of the consistently averaged quantity $\lim_{r\to\infty} \chip^{\rm (av)}$. Gray contours indicate the posterior distribution of GW190521 (68\% and 90\%  levels). Samples have been backpropagated to $f_{\rm GW}=0$ Hz ($r\to \infty$) to model how the source formed. %
 }
\label{fig:chipeff}
\end{figure*}

\subsection{Elliptical arcs}

Let us assume for simplicity that first-generation (1g) BH binaries have uniformly distributed spin magnitudes $\chi_i\in [0,\chimax]$, while their mass ratio $q$ has a power-law distribution $p(q)\propto q^\beta$. Large, positive values of $\beta$ indicate a scenario where BHs pair selectively with companions of similar masses, as expected in mass-segregated clusters. The case $\beta=0$ models a random pairing process, and it is tentatively supported by current GW observations~\cite{Abbott:2020gyp}.  We compute the effective spins for 1g+1g mergers by assuming the spin directions to be isotropically distributed, as expected in dynamical formation channels. 

From this population of 1g+1g mergers, we estimate the spin of the second-generation (2g) merger remnants using fits to numerical relativity simulations~\cite{Hofmann:2016yih}. This yields the probability distribution $p(\chi_r)$ of 2g BHs. To construct the population of 1g+2g and 2g+2g mergers, the 2g BHs [with spin distribution $p(\chi_r)$] are paired with 1g and 2g BHs [with spin distributions $U(0, \chimax)$ and $p(\chi_r)$] assuming the pairing probability $p(q)\propto q^\beta$. In Fig.~\ref{fig:chipeff} we show the three  populations, 1g+1g, 1g+2g, and 2g+2g, where we contrast $\chieff$ against  both the heuristic  expression $\chip^{(\rm heu)}$ [Eq.~(\ref{eq:chiph})] and the asymptotic limit of the fully averaged expression $\chip^{(\rm av)}$ [Eq.~(\ref{ellipE})]. 

We find that 1g+1g BH binaries are confined to a region near the origin of the $(\chieff,\, \chip)$ plane. Interestingly, the populations involving 2g BHs cluster in arc-shaped structures extending to regions of the plane where $\chip \gtrsim 0.5$ and $|\chieff|\gtrsim 0.5$. 

The leftmost panels of Fig.~\ref{fig:chipeff} show a case where BH spins at birth are low ($\chimax=0.1$) and pairing is highly selective ($\beta=5$). 
In this case, the 1g+2g and 2g+2g spin distributions are well separated from their 1g+1g progenitors, indicating that  future GW measurements with accuracies of $\sim 10\%$ on $\chieff$ and $\chip$ will allow us to confidently pinpoint their hierarchical generation. 

To better understand this analytically, let us analyze the limit where $\chimax \ll1$ and $\beta\gg 1$. %
Because pairing is selective, most sources will have $q\sim 1$.

 For the 1g+2g population, one has $\chi_2\sim 0$ (because 1g spins  are assumed to be low) and $\chi_1\sim0.7$ (because the 2g BH is the remnant of a previous merger). The condition $\chi_2\ll \chi_1$ implies that  $\chi_{\rm p}^{\rm (heu)}\simeq \lim_{r\to\infty} \chip^{\rm (av)}$  \cite{Gerosa:2020aiw}, thus explaining why the two orange arcs in the left panels of Fig.~\ref{fig:chipeff} are very similar to each other. From the definitions \eqref{eq:chieff} and \eqref{eq:chiph} it follows that $\chieff\simeq \chi_1\cos\theta_1/2$ and $\chip\simeq \chi_1\sin\theta_1$, and therefore
\begin{equation}
\left(\frac{2 }{0.7} \chieff\right)^2 + \left(\frac{\chip}{0.7}\right)^2 \simeq 1\,,
\label{ellipse}
\end{equation}
which is the equation of an ellipse. The effective spins are thus limited by  $-0.35\lesssim \chieff\lesssim 0.35$ and $0\lesssim \chip \lesssim 0.7$. 

For the 2g+2g case, one has $q\sim 1$ and $\chi_1\sim\chi_2\sim 0.7$. This yields the relation
\begin{equation}
\left(\frac{2}{0.7}\chieff+\cos\vartheta \right)^2 + \left(\frac{\chip^{\rm (heu)}}{0.7}\right)^2 \simeq 1\,,
\label{shiftedellipse}
\end{equation}
where $\cos\vartheta$ is a random number distributed uniformly in $[-1,1]$. The additional term compared to Eq.~(\ref{ellipse}) has the effect of splitting the ellipse and shifting it horizontally in both directions, resulting in the double-arc blue structures observed in Fig.~\ref{fig:chipeff}. Equation~(\ref{shiftedellipse}) also implies that $\chip^{\rm (heu)}\lesssim 0.7$.  However this is an artificial limit introduced by the heuristic definition of $\chip$, as evidenced by the sharp truncation of the blue arcs in the top panels of Fig.~\ref{fig:chipeff}. We can estimate the upper bound of the fully averaged precession parameter  $\chip^{\rm (av)}$ by imposing $\chieff\sim 0$, which in this case corresponds to setting $\cos\theta_1\sim -\cos\theta_2$ and thus $\chi_{1\perp}\sim\chi_{2\perp}$ in Eq.~(\ref{ellipE}), with the result
\begin{equation}
\lim_{r\to\infty} \chip^{\rm (av)}\lesssim \frac{4\times 0.7}{\pi} \sqrt{1-\cos^2\vartheta}\lesssim 0.9\,.
\end{equation}
This is consistent with the extended blue arcs in the bottom-left panel
of Fig.~\ref{fig:chipeff}.

The trends that we just described remain valid, although in a more approximate fashion,  for generic values of $\beta$ and $\chimax$.
For uniform pairing ($\beta=0$; rightmost panels in Fig.~\ref{fig:chipeff}), the 2g spins follow a broader distribution extending from $\ssim 0$ to $\ssim 0.7$. This increases the area covered by the arcs in Fig.~\ref{fig:chipeff}. Small-spin 2g mergers end up populating the region where $\chieff$ and $\chip$ are both small, leading to some overlap with the 1g+1g distribution.  If instead the 1g spins are larger ($\chimax=0.5$; middle panels in Fig.~\ref{fig:chipeff}), the 1g+1g spins end up spanning a larger region in the $(\chieff,\, \chip)$ plane, but the 2g spins remain peaked at large values.

\subsection{Backpropagating GW190521}

At present, the most promising candidate of hierarchical origin is GW190521~\cite{Abbott:2020tfl}. This conclusion is largely driven by its masses, at least one of which lies in the pair-instability gap. The ``effective-spin arcs'' that we just explored allow us to cross-check whether this interpretation is also compatible with its spin properties. The measured values of $
\chieff$ and $\chip$ for GW190521 are indicated in Fig.~\ref{fig:chipeff} with gray contours. In the posterior distribution samples provided in Ref.~\cite{Abbott:2020niy}, the spin directions are provided at the arbitrary reference frequency $f_{\rm GW}=11$ Hz. We backpropagate the spin distributions to $f_{\rm GW}=0$ Hz using the formalism and the code of Refs.~\cite{Kesden:2014sla,Gerosa:2015tea,Gerosa:2016sys}. 
This is because, much like any other astrophysical model, our populations describe BH binaries at formation, not at detection, so it would be wrong to use the LIGO posterior samples at face value. 
To the best of our knowledge, this is the first time that samples from a real GW event have been backpropagated from detection to formation.%

GW190521 is more easily accommodated by the 1g+2g and 2g+2g populations for all values of $\chimax$ and $\beta$ explored in Fig.~\ref{fig:chipeff}.  %
To quantify this statement, we compute the likelihood that GW190521 belongs to generation $g$ as follows:
\begin{align}
&P_g(\chimax,\beta, | d) = \notag\\
& \int d\chieff \, d\chip \frac{p(\chieff,\chip | d)}{\pi(\chieff,\chip)} p_{g}(\chieff,\chip| \chimax,\beta)
\label{eq:gen lkl}
\end{align}
where $g$=\{1g+1g, 1g+2g, 2g+2g\}, $p(\chieff,\chip | d)$ is the LIGO posterior for GW190521, $\pi(\chieff,\chip)$ is the LIGO prior, and $p_g(\chieff,\chip| \chimax,\beta)$ is the probability distribution of $(\chieff,\,\chi_p)$ for given $(\chimax,\,\beta)$. Note that in the expression above we are implicitly neglecting selection effects~\cite{Mandel:2018mve,Vitale:2020aaz}, which are largely irrelevant in this case  because we are  integrating only over the spin parameters~\cite{Ng:2018neg}. Since the event priors and posteriors are provided as discrete samples, the integral in Eq.~(\ref{eq:gen lkl}) is computed as a Monte Carlo summation with $p_g$ and $\pi$ evaluated on the posterior samples. The prior values are estimated with a bounded kernel density fit to the LIGO prior samples.

The likelihood ratios
\begin{align}\label{eq:l_ratio1}
\mathcal{L}_{\rm 1g+2g} = \frac{P_{\rm 1g+2g}(\chimax,\beta, | d)}{P_{\rm 1g+1g}(\chimax,\beta, | d)}
\end{align}
and
\begin{align}\label{eq:l_ratio2}
\mathcal{L}_{\rm 2g+2g} = \frac{P_{\rm 2g+2g}(\chimax,\beta, | d)}{P_{\rm 1g+1g}(\chimax,\beta, | d)}
\end{align}
are given in Table~\ref{tab:l_ratio} for all values of $\chimax$ and $\beta$ explored in Fig.~\ref{fig:chipeff}. For small 1g spins and selective pairing ($\chimax=0.1$ and $\beta=5$), large precessing spins of GW190521 can only be explained by 2g populations. On the other hand, if $\chimax=0.5$, 1g populations have extended support to large $(\chieff,\,\chi_p)$, leading to a reduction in  $\mathcal{L}_{\rm 1g+2g} $ and $\mathcal{L}_{\rm 2g+2g}$. 
In the limit $\chimax\to1$, which allows for the possibility that the large-spin BHs observed in high mass X-ray binaries~\cite{Miller:2014aaa} may contribute to multiple-generation mergers, one has $\mathcal{L}_{\rm 1g+2g}\to 1.1$ ($\mathcal{L}_{\rm 2g+2g}\to 1.1$) if $\chi_p=\chi_p^{\rm (heu)}$ or $\mathcal{L}_{\rm 1g+2g}\to 1.3$ ($\mathcal{L}_{\rm 2g+2g}\to 1.4$) if $\chi_p=\chi_p^{\rm (avg)}$ , blurring the lines between the spin distribution of 1g+1g mergers and mergers involving 2g BHs.
Similarly, random pairing at $\beta=0$ also leads to smaller 2g spins, and hence relatively less support at large spins, again reducing  $\mathcal{L}_{\rm 1g+2g} $ and $\mathcal{L}_{\rm 2g+2g}$. While large values of $\chimax$ and small values of $\beta$ reduce the likelihood ratio, they might also diminish the population of 2g BHs. This is because the larger recoils experienced by remnants of 1g+1g mergers will make it harder for clusters to retain them.

In Table~\ref{tab:l_ratio} we did not include a prior ratio, which is model dependent.

\begin{table}

\begin{tabular}{C||CC|CC}
\hline
&\multicolumn{2}{C|}{\chi_p=\chi_p^{\rm (av)}} & \multicolumn{2}{C}{\chi_p=\chi_p^{\rm (heu)}}\\
\!\!(\chimax,\,\beta)\!\! &  \!\!{\mathcal L}_{\rm 1g+2g}
\!\!\!\!  & {\mathcal L}_{\rm 2g+2g} \!\!& \!\!{\mathcal L}_{\rm 1g+2g}\!\!\!\!  & \!\!{\mathcal L}_{\rm 2g+2g}\!\!\!\!\\
\hline
(0.1,\,5) & 5.3 & 8.9& 6.1 &6.3\\
(0.5,\,5) & 3.5& 5.1&3.4&3.6\\
(0.1,\,0) & 3.8&4.7 & 4.5& 4.8\\
\hline
\end{tabular}
  \caption{Likelihood ratios for GW190521 when comparing 1g+2g and 2g+2g origins against 1g+1g origin.}
\label{tab:l_ratio}
\end{table}

\section{The parents of hierarchical black-hole mergers}
\label{parents}

Constraints on hierarchical mergers would become more informative if GW data were to provide accurate measurements of the individual spins of the merging BHs---not only of some effective combination thereof. In this case, one can reconstruct not only the generation of the observed events but also the properties of its parents.  

\subsection{Constraints on the remnant black-hole spin}

Suppose that we have measured the dimensionless spin $\chirem$ of one binary component. This is likely to be the spin of the heavier BH in the observed binary (i.e. $\chirem=\chi_1$), although this is not required for the argument presented here to be valid. Assuming that the measured BH with spin $\chirem$ is the remnant of a previous merger, we now wish to infer the properties of its parents.

Consider a parent binary with masses $M_{1,2}$, spins $\boldsymbol{\xi}_{1,2}$ and mass ratio $Q=M_2/M_1\leq 1$ (recall that $m_{1,2}$, $\boldsymbol{\chi}_{1,2}$ and $q=m_2/m_1$ denote the corresponding quantities for the observed binary; cf. Fig.~\ref{cartoonparents}). Using the numerical-relativity fits of %
Ref.~\cite{Barausse:2009uz},
$\chirem$ is given by %
\begin{align}
&\chirem=\min\Bigg\{1, \Bigg|\f{\vec{\chiparent}_1+\qp^2  \vec{\chiparent}_2 }{(1+\qp)^2} + \eta \vec{\hat L} 
\Bigg[ 2\sqrt{3} + t_2 \eta + t_3 \eta^2 
\notag \\ & 
+\!s_4 \left(\!\f{ \vec{ \chiparent}_1+\qp^2  \vec{ \chiparent}_2}{1+\qp^2} \!\right)^2 \!\!\!
+\!(s_5 \eta + t_0 +2)\! \left(\!\f{\vec{\chiparent}_1+\qp^2  \vec{\chiparent}_2}{1+\qp^2} \!\right)\!\cdot\!\vec{\hat L}\Bigg]
\Bigg|\Bigg\} \label{finalspin}\,,
\end{align}
where $\eta=Q/(1+Q)^2$ %
and $\vec{L}$ denotes the angular momentum of the parent binary. The fitting coefficients are 
 $t_0=-2.8904$, $t_2=-3.51712$, $t_3=2.5763$, $s_4=-0.1229$, and $s_5=0.4537$~\cite{Barausse:2009uz}.
We use the fits of Ref.~\cite{Barausse:2009uz} as opposed to newer fits like Refs.~\cite{Hofmann:2016yih,Varma:2018aht} to keep the analytical treatment of the problem tractable. This has a negligible effect on our analysis, because the systematic error in the fits is smaller than typical statistical errors in the measurement of $\chi_r$.

When both spins of the parent binary are either aligned ($\vec{\hat L}\cdot\vec{\hat{\xi}}_{1}  = \vec{\hat L}\cdot\vec{\hat{\xi}}_{2}= +1$, 
henceforth denoted by $\uparrow\uparrow$) or antialigned ($\vec{\hat L}\cdot\vec{\hat{\xi}}_{1}  = \vec{\hat L}\cdot\vec{\hat{\xi}}_{2}= -1$ denoted by $\downarrow\downarrow$) relative to the orbital angular momentum,  $\chirem$  is given by 
\begin{align}\label{remMinMax}
\chirem^{\uparrow\uparrow/\downarrow\downarrow}& = \Bigg|2\sqrt{3}\eta + t_2 \eta^2 + t_3 \eta^3 + \f{s_4\eta}{(1-2\eta)^2} \chit^2\nn\\& \pm\left[1+\eta \f{(s_5 \eta + t_0 +2)}{1-2 \eta}\right]\chit\Bigg|\,,
\end{align}
where 
\be
\chit=\f{\chionep +\qp^2 \chitwop }{(1+\qp)^2}\,.
\label{defchit}
\ee
The plus and minus signs correspond to aligned ($\uparrow\uparrow$) and antialigned ($\downarrow\downarrow$) spins, respectively.  We dropped the minimum operation that was present in Eq.~(\ref{finalspin}) and, consequently, our inversion is valid only for $\chirem<1$.

In most of the three-dimensional parameter space $(\qp,\,\chionep,\,\chitwop)$, the remnant spin $\chirem$ is maximal when the binary spins are aligned ($\chirem=\chirem^{\uparrow\uparrow}$) and it is minimal when they are antialigned ($\chirem=\chirem^{\downarrow\downarrow}$). In the corner of the parameter space with $\qp/(1+\qp)^2\lesssim 0.28 \chit$ , %
the minimum $\chirem$ is instead obtained when the primary BH is antialigned and the secondary BH is aligned, i.e. %
$\vec{\hat L}\cdot\vec{\hat{\xi}}_{1}  = -\vec{\hat L}\cdot\vec{\hat{\xi}}_{2}= -1$
(denoted by $\downarrow\uparrow$). The value of $\chirem^{\downarrow\uparrow}$ can be obtained by replacing  $\chit\to ({\chionep -\qp^2\chitwop })/({1+\qp)^2}$ 
 in the expression for $\chirem^{\downarrow\downarrow}$. %
Because 
\be
\frac{\chionep -\qp^2\chitwop}{(1+\qp)^2}=\chit+\mathcal{O}(\qp^2)
\ee
and the condition $\chirem^{\downarrow\uparrow}<\chirem^{\downarrow\downarrow}$ requires unequal mass binaries, in the following we approximate the minimum of $\chirem$ with $\chirem^{\downarrow\downarrow}$ for all values of $\qp$, $\chionep$, and $\chitwop$. The error introduced by this approximation is $\Delta\chirem\lesssim 0.07$, and it is largest for the extremal spins $\chionep\sim 1$ and $Q\sim 0.26$.

\begin{figure}[t]
  \includegraphics[width=\columnwidth]{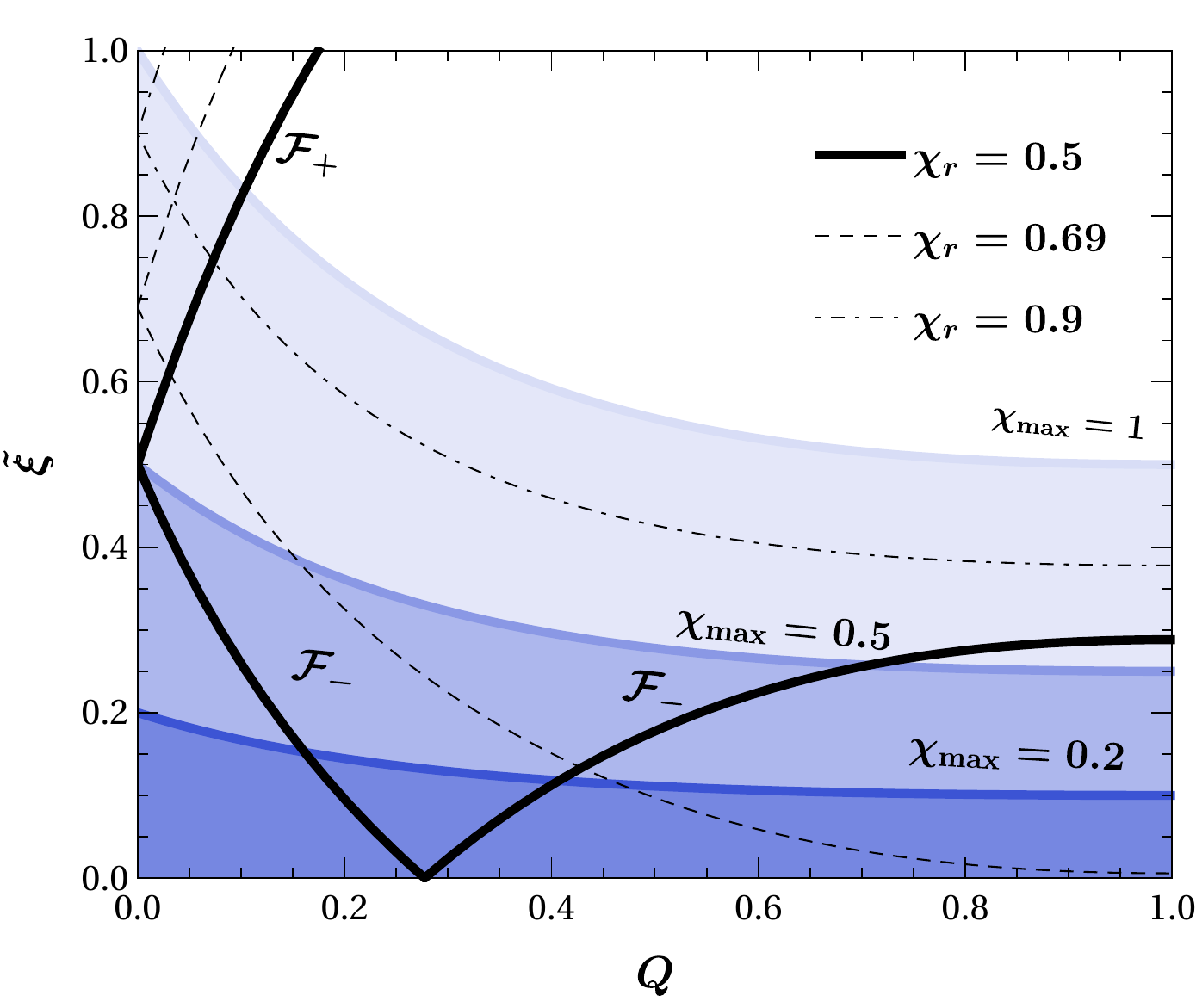}
  \caption{Bounds on the mass ratio $Q$ and spin combination $\chit$ of the parent binary producing a remnant with spins $\chirem=0.5$ (solid black line), $\chirem=0.69$ (thin dashed line), and $\chirem=0.9$ (thin dash-dotted line). The lower and upper bounds are given by the functions $\mathcal{F}_\pm$ defined in Eq.~(\ref{eq:bounds_chi}). For clarity, in this figure we explicitly label these functions only when $\chirem=0.5$. Blue lines represent further upper limits that can be imposed if the parent BH spins are bounded by $\chiparent_{1,2}<\chimax$ with $\chimax=0.2$, $0.5$ and $1$ (in different shades of blue): cf. Eq.~(\ref{eq:bounds_chimax}). For given values of $\chirem$ and $\chimax$, the allowed region for the progenitors lies within the wedges above the black curves corresponding to the given $\chirem$, and below the blue curves corresponding to the given $\chimax$.
}
\label{fig:chit}
\end{figure}

\begin{figure}[t]
  \includegraphics[width=\columnwidth]{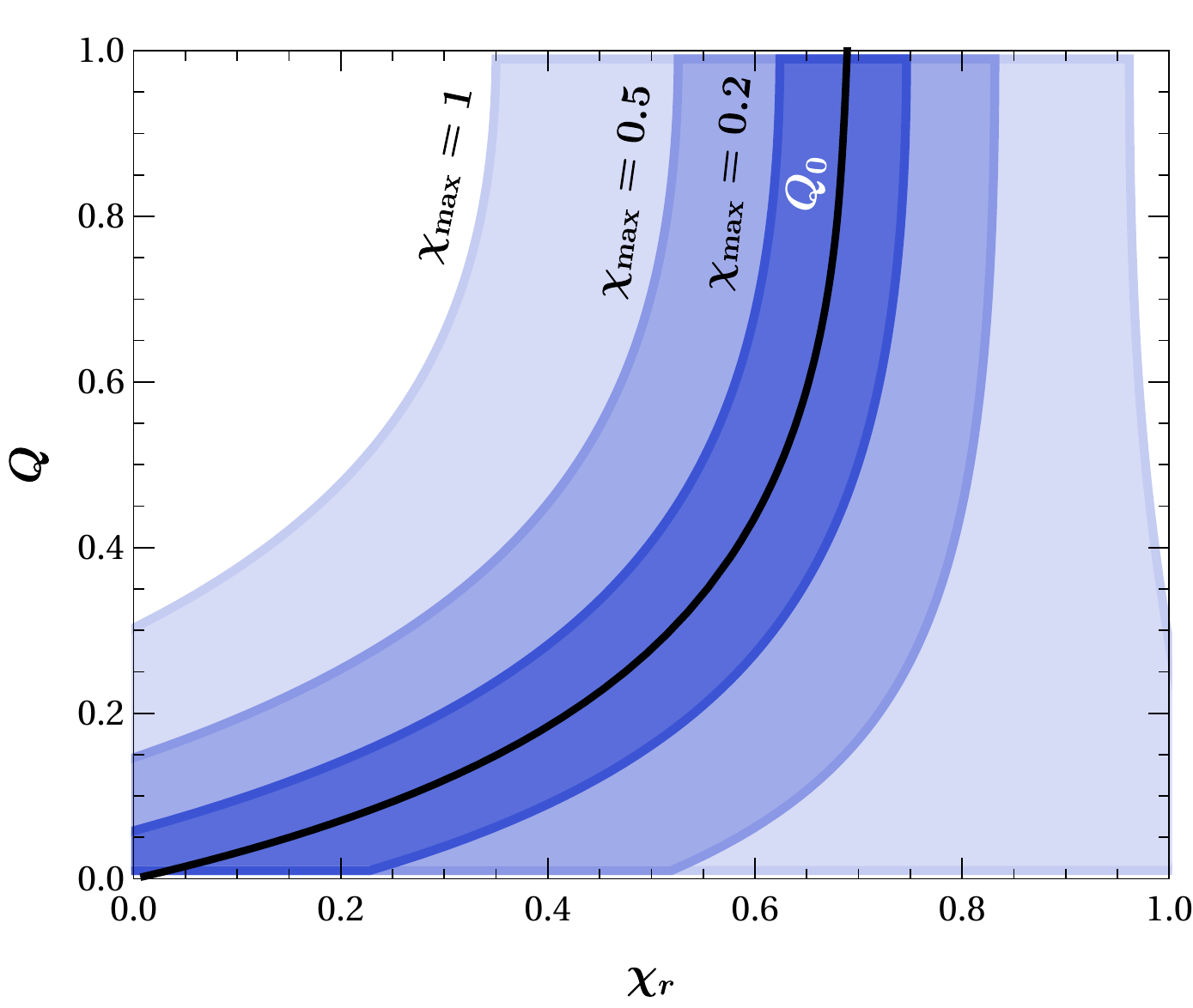}
  \caption{Allowed range of the parent binary mass ratio $\qp$ for a given value of the observed spin $\chirem$. A single value $\qp=\qp_0$ (solid black curve) is allowed if both parents are nonspinning, i.e. $\chimax=0$. As $\chimax$ increases, the allowed range of $\qp$ for a given $\chirem$ (shown in different shades of blue) widens. }
\label{fig:q0}
\end{figure}

\subsection{Inferring the parents' mass ratio from the remnant spin}
\label{criticalq}

Our next goal is to find the spin combinations that yield a given $\chirem$. We can invert the relation
\be\label{eq:bounds_rem}
\chirem^{\downarrow\downarrow}(\qp,\chit) \le\chirem\le\chirem^{\uparrow\uparrow}(\qp,\chit)
\ee
to obtain the $\chit$ needed to form a remnant with spin $\chirem$:
\be\label{eq:bounds_chi}
{\mathcal F}_-(\chirem, \qp)\le\chit\le {\mathcal F}_+(\chirem, \qp)\,,
\ee
where the functions ${\mathcal F}_-(\chirem, \qp)$ and  ${\mathcal F}_+(\chirem, \qp)$ are obtained by inverting Eq.~(\ref{remMinMax}) %
and are shown in Fig.~\ref{fig:chit}. %
The detection of a BH with spins $\chirem$ restricts the progenitor's properties to a specific wedge of the ($Q,\chit$) plane. 

There is a special value of the mass ratio $\qp_0$ for which ${\mathcal F}_-(\chirem, \qp_0)=0$. This case corresponds to the merger of nonspinning BHs ($\chionep=\chitwop=0$). From Eq.~(\ref{finalspin}) 
one gets
\be\label{eq:q0}
\chirem=\eta_0\left(2\sqrt{3} + t_2 \eta_0 + t_3 \eta_0^2\right)\,,
\ee
where $\eta_0=\qp_0/(1+\qp_0)^2$. For instance, for $\chirem=0.5$   one has $\qp_0=0.28$. If the parent binary had a mass ratio $\qp_0$, there would be no restriction on the component spin magnitudes of the progenitors that could have formed the observed   remnant.  For all other values of $\qp\neq \qp_0$, the parents must have had nonzero spin magnitudes. 

When the remnant spin reaches a value $\chirem\simeq 0.69$, we find that $\qp_0=1$. Equal-mass parents can form this remnant irrespective of their spin magnitudes, but larger BH spins are required  if $\qp$ is smaller. 
If $\chirem\gtrsim0.69$,  the equation ${\mathcal F}_-(\chirem, \qp_0)=0$ does not admit solutions with $Q_0\in [0,1]$. Nonspinning BHs cannot possibly form such a remnant. If interpreted as a hierarchical merger, a putative GW observation of a BH with $\chirem \gtrsim0.69$ would indicate that the BHs of the previous generation were also spinning. %

One might want to further impose an upper bound on the spin magnitudes of the previous-generation BHs, say $\xi_{1,2}\leq\chimax$. If the parent's binary consists of  1g BHs, the parameter $\chimax$ has the clear physical interpretation of the largest BH spin resulting from stellar collapse. This additional condition translates  to 
\be\label{eq:bounds_chimax}
\chit \le \f{1+\qp^2}{(1+\qp)^2} \,\chimax \,,
\ee
and further restricts the allowed region in the ($\qp,\chit$) plane. The upper bounds corresponding to selected values of $\chimax$ are shown in Fig.~\ref{fig:chit} as solid colored lines. For instance, if $\chirem=0.5$ and $\chimax=0.1$, the parent's mass ratio must satisfy $0.22 \le \qp \le 0.34$. 

This point is further explored in Fig.~\ref{fig:q0}. For a given value of $\chirem$, there is a finite range of allowed mass ratios $Q$. The width of this range depends on the largest spin magnitude of the progenitor $\chimax$. In particular,  it shrinks to zero for $\chimax=0$ (corresponding to the condition $Q=Q_0$) and widens progressively if larger progenitor spins are allowed.

\begin{figure}[t]
  \includegraphics[width=\columnwidth]{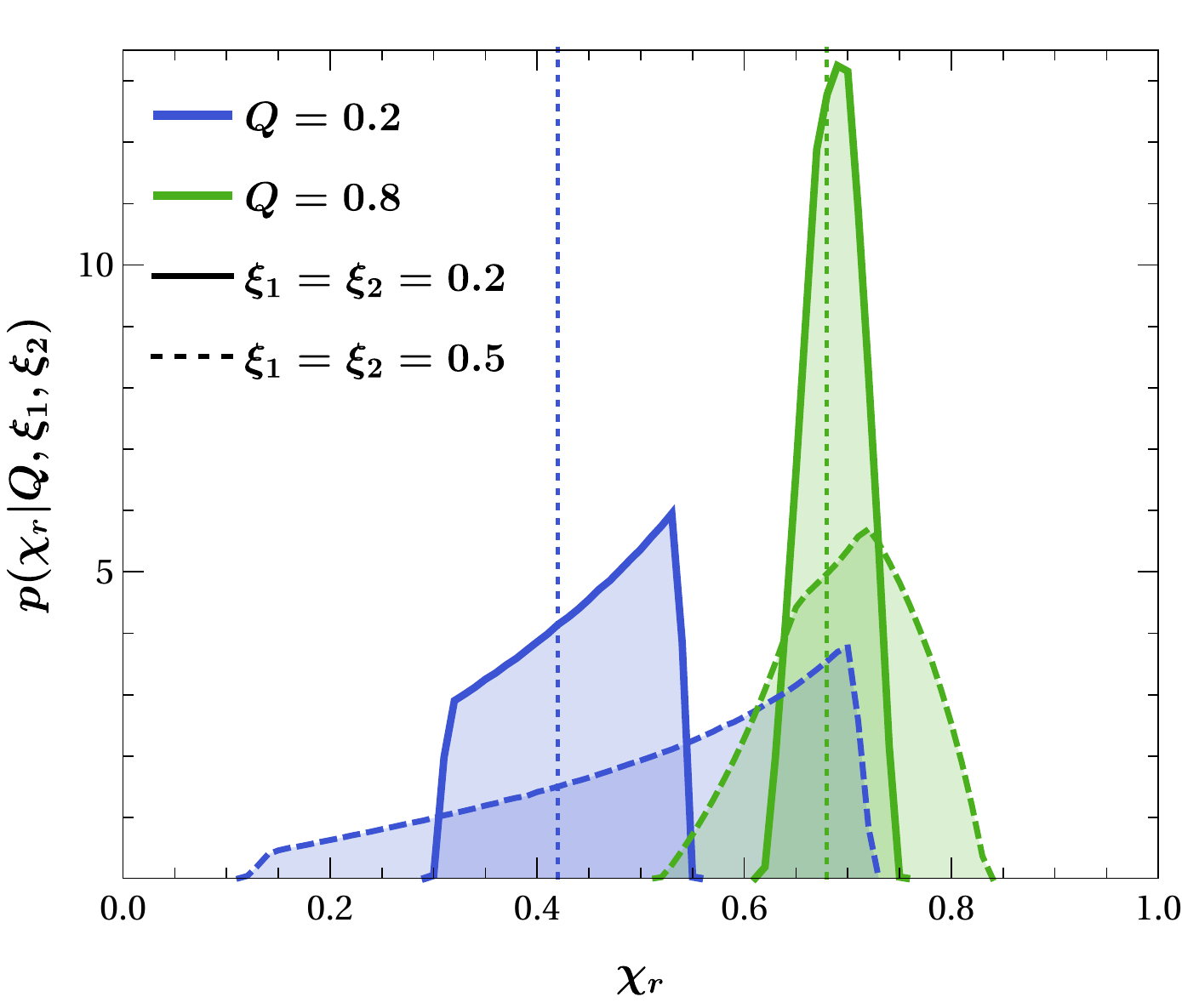}
  \caption{Probability density function of the primary spin in a putative 1g+2g GW event as a function of the mass ratio $\qp$ and spin magnitude $\xi_{1,2}$ of the parent binary. 
     Blue (green) curves refer to $\qp=0.2$ ($\qp=0.8$), and solid (dashed) lines refer to $\chionep=\chitwop=0.2$ ($\chionep=\chitwop=0.5$). Vertical lines show the remnant spin values for the simpler case where both parents have zero spin, $\chionep=\chitwop=0$. %
}
\label{fig:chirpdf}
\end{figure}

\begin{figure*}[t]
  \includegraphics[width=0.9\textwidth]{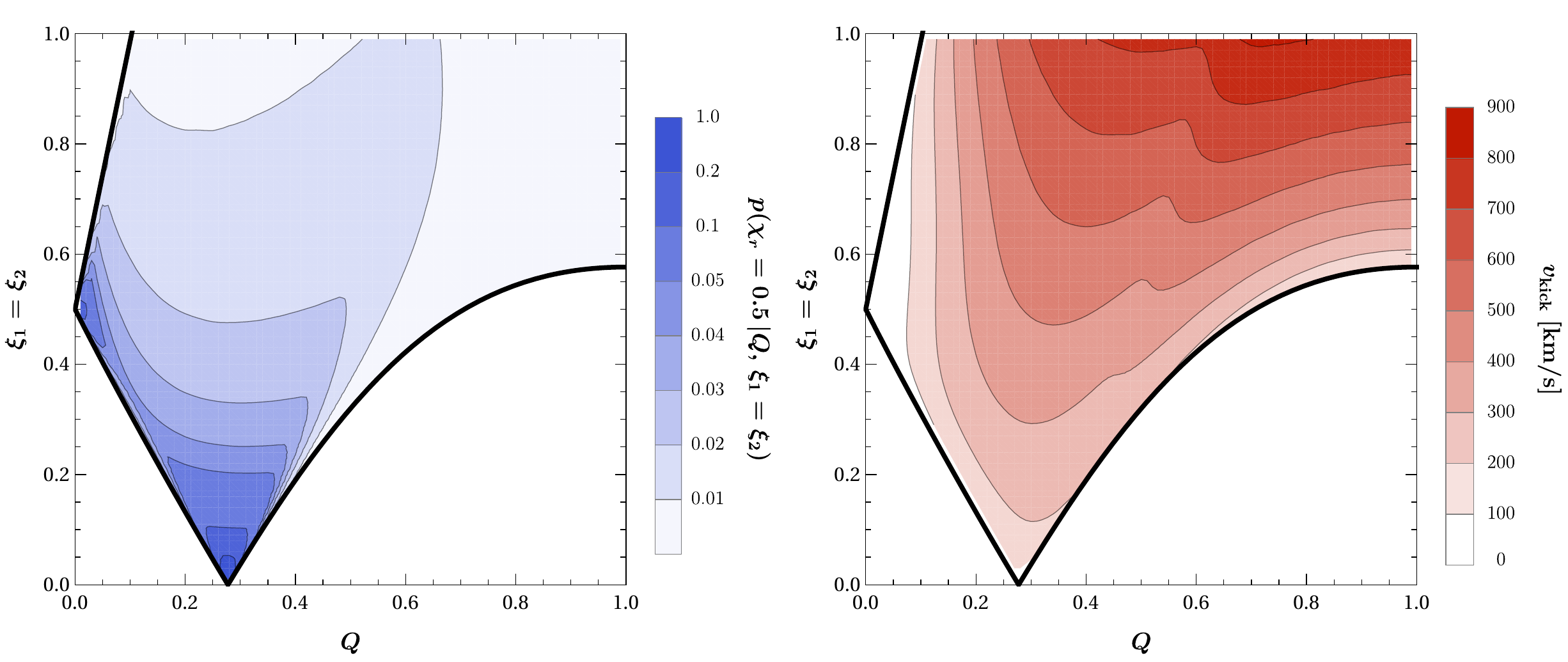}
  \caption{Left panel: robability $p(\chirem=0.5 | \qp, \chionep=\chitwop )$ (in arbitrary units) as a function of the parent's mass ratio $\qp$ and spins $\chionep=\chitwop$ (here taken to be equal to each other for simplicity), assuming isotropic spin directions. Right panel: kick velocity $v_{\rm kick}$ imparted to BHs with spin $\chirem=0.5$ as a function of the mass ratio $\qp$ and spins $\chionep=\chitwop$ of the merging binary. Areas of high probability (dark regions in the left panel) correlate with areas of low kick (light regions in the right panel). The white areas beyond the solid black lines represent forbidden regions where a remnant with $\chirem=0.5$ cannot be formed. 
 }
\label{fig:pcq}
\end{figure*}

\begin{figure}[t]
  \includegraphics[width=\columnwidth]{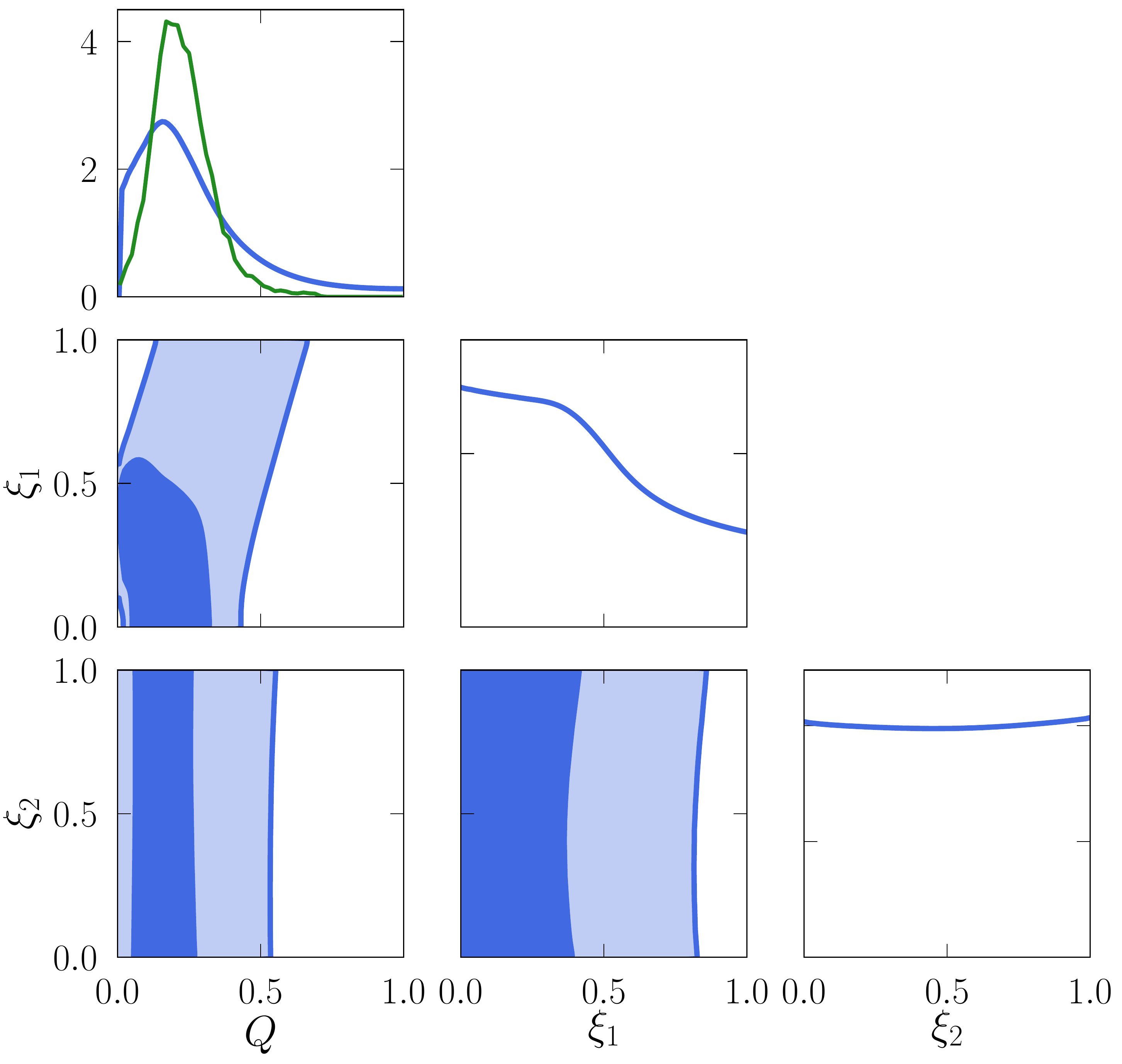}
  \caption{Assuming that the primary BH in GW190412 is of hierarchical nature, we show (in blue) the posterior distribution of the mass ratio $\qp$ and spin magnitudes $\xi_{1,2}$ of its parents.  Contours indicate the $50$\% and $90$\% confidence intervals.  The medians and 90\% credible intervals of the marginalized distributions are $\qp=0.21_{-0.16}^{+0.32}$, $\xi_1=0.37^{+0.45}_{-0.30}$, and $\xi_2=0.50_{-0.40}^{+0.40}$. The green line shows the posterior of $Q$ under the assumption that $\xi_{1,2}=0$, with median and $90\%$ confidence intervals $\qp=0.22_{-0.11}^{+0.14}$.}
\label{fig:corner}
\end{figure}

\subsection{Not all parents are equally likely}
\label{parlikely}

The exclusion regions on the parents' properties presented so far were obtained by requiring that there exists some orientation of the progenitor spins $\vec{\hat{\xi}}_{1,2}$ that can produce a given remnant. The distribution  $p(\chirem| \qp,\,\chionep,\,\chitwop)$ is shown in Fig.~\ref{fig:chirpdf}, where isotropic spin directions are assumed, as expected for dynamical formation channels that may produce hierarchical mergers.
 
As $\qp\to1$, the probability is narrowly peaked at around $0.7$. The location of the peak value of $\chirem$ decreases at smaller mass ratios. Larger progenitor spins also lead to broader distributions of remnant spins (see e.g.~\cite{Berti:2008af,Gerosa:2021hsc}). 

Using the inversion formalism described above, we can now explore which of the possible parents are more or less likely. The left panel of  Fig.~\ref{fig:pcq} illustrates how the probability of forming a BH of spin $\chirem=0.5$ depends on the progenitors' spins and mass ratio. We have fixed $\chionep=\chitwop$ to reduce the number of parameters for illustrative purposes, but the argument can be made more general.  
The solid black lines represent the bounds derived from Eq.~(\ref{eq:bounds_chi}). 
There are two distinct regions of parameter space where the probability is highest: 
\begin{enumerate}[label=(\arabic*)]
\item When the mass ratio is close to the critical value $\qp\simeq \qp_0$ and the spins are small, i.e. $\chiparent_{1,2}\simeq 0$. 
\item
When $\qp\simeq 0$ and $\chionep\simeq \chirem$, corresponding to the trivial case of a very small body merging into a larger Kerr BH with spin $\chirem$. 
\end{enumerate} 
The spin orientations lose meaning in both the zero-spin (case 1) and test-particle (case 2) limits, which implies a higher concentration of possible remnants.

Crucially, these are the very same limits that minimize the kick imparted to the BH remnant. This is illustrated in the right panel of Fig.~\ref{fig:pcq}, where we use the kick fitting formula of Ref.~\cite{Gerosa:2016sys}, selecting the spin directions that can produce the targeted $\chirem$. The kick is suppressed in both of the ``special'' regions near $Q\simeq0$ and $\chionep=\chitwop\simeq 0$. Parents with these properties are not only more likely to form a remnant with the targeted properties, but also more likely to form a remnant that is retained in its astrophysical host---a prerequisite condition in the scenario depicted here, where we are observing the GWs emitted by the subsequent merger. %

\subsection{Application to GW190412}

The arguments above require that the spin of one BH binary component can be reliably measured from the data. As explored in Sec.~\ref{chieffchipplane}, we often have access only to the effective spins. For $q\ll1$, the spin of the secondary is negligible and one can reconstruct the spin of the primary BH as $\chi_1\simeq \sqrt{\chieff^2+\chip^2}$.

Although most of the binaries detected to date are compatible with having comparable masses, there are two notable exceptions: GW190412 with $q=0.28_{-0.07}^{+0.13}$, and GW190814 with $q=0.112^{+0.008}_{-0.008}$~\cite{Abbott:2020niy}.  %
In particular, GW190412 was shown to be compatible with a hierarchical merger~\cite{Gerosa:2020bjb} and was interpreted as such in both cluster~\cite{Rodriguez:2020viw} and AGN~\cite{Tagawa:2020dxe} formation models. Using the spin inversion formalism derived in this paper and the LIGO measurement $\chi_1= 0.44_{-0.22}^{+0.16}$~\cite{Abbott:2020niy}, can we infer the properties of the parents of GW190412?  %

Given the event data $d$, we calculate
\be\label{eq:posterior}
p(\qp, \chionep, \chitwop|d) \propto \pi(\qp, \chionep, \chitwop)  \int p(\chirem | \qp, \chionep, \chitwop)\, p(\chirem | d) \, d\chirem\,,
\ee
where $p(\chi_1 | d)$ denotes the LIGO posterior for GW190412, $p(\chirem | \qp, \chionep, \chitwop)$  is the probability distribution introduced in  Sec.~\ref{parlikely} (where we assumed isotropic orientations), and $\pi(\qp, \chionep, \chitwop)$ is a prior, which we assume to be flat in $\qp\in [0,1]$ and $\chiparent_{1,2}\in[0,1]$. %
We sample  $p(\qp, \chionep, \chitwop|d) $ using Markov-chain Monte Carlo  \cite{2013PASP..125..306F}. 

Figure~\ref{fig:corner} shows the resulting constraints on the parents of GW190412's primary BH. In particular, we infer that the  progenitor binary must have had a relatively small ratio $\qp=0.21_{-0.16}^{+0.32}$. %
This follows mainly from the small value of $\chirem \approx 0.44$. As illustrated in Fig.~\ref{fig:pcq}, the regions around $\qp\simeq \qp_0,\,\chiparent_{1,2}\simeq 0$ and $\qp\simeq 0,\,\chionep \simeq \chirem$ are favored. Similar features can also be seen in Fig.~\ref{fig:corner}, although these are somewhat washed out by measurement  errors on $\chirem$.
Small values of $\chionep$ are favored, while $\chitwop$ cannot be inferred. If the mass ratio $\qp$ is small, one has $\chit\approx\chionep$, and the likelihood becomes largely independent of $\chitwop$. %

Now that we have (samples of) the parent binary's properties, we can estimate the energy $E_{\rm rad}$ that was dissipated in GWs during the earlier merger that formed the primary component of GW190412. For each sample in the three-dimensional parameter space $(Q,\xi_1,\xi_2)$ of Fig.~\ref{fig:corner}, we solve for the spin directions $\hat{\vec{\xi}}_{1,2}$ that correspond to a given remnant spin $\chirem$ in the LIGO posterior. We then plug the resulting values of $(Q,\vec{\xi}_1,\vec{\xi}_2)$ into the numerical-relativity fit of Ref.~\cite{Barausse:2012qz} to estimate $E_{\rm rad}$.  When combined with the samples of $m_1=30.1^{+4.6}_{-5.3}$~\cite{Abbott:2020niy} provided by LIGO, this allows us to reconstruct the masses
\begin{align}
M_1 &= \frac{m_1}{1+Q}(1-E_{\rm rad})\,,\ \\
M_2 &= \frac{Q m_1}{1+Q}(1-E_{\rm rad}) \
\end{align}
of the parent BHs.

\begin{figure}[t]
  \includegraphics[width=\columnwidth]{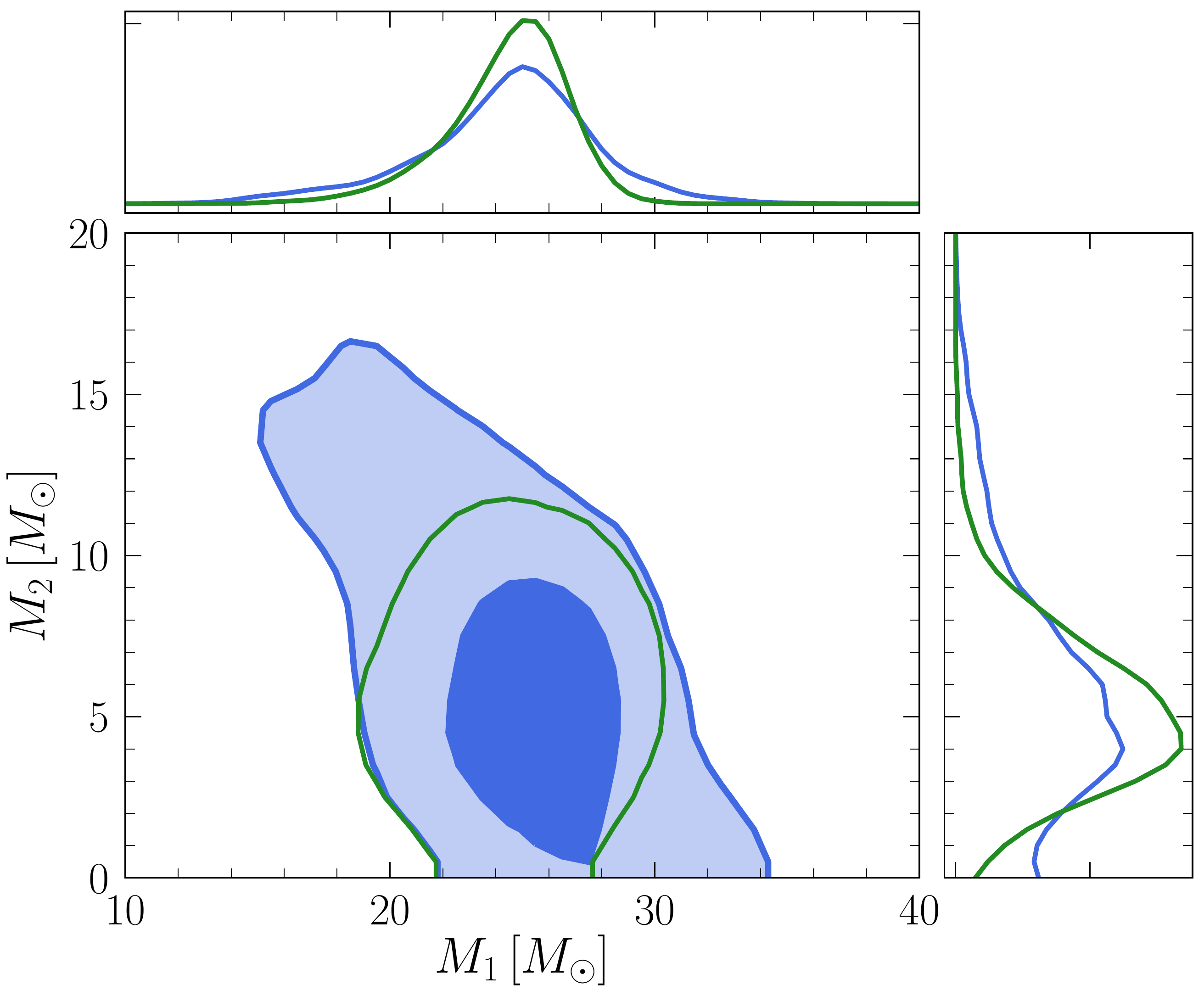}
  \caption{Masses $M_1$ and $M_2$ of the parents of the primary component in GW190412 (blue) under the uniform-spin priors. Contours indicate the $50$\% and $90$\% confidence intervals.  The medians and 90\% credible intervals of the marginalized distributions are $M_1=  25.24^{+3.55}_{-4.83} \Msun$ and $M_2=5.61^{+5.67}_{-3.98}\Msun$. The green lines show $90$\% confidence intervals when the progenitors are nonspinning: $M_1=25.24^{+2.35}_{-3.35}\Msun$ and $M_2=5.44^{+3.41}_{-2.80}\Msun$.
  } 
\label{fig:mcorner}
\bigskip\medskip
  \includegraphics[width=\columnwidth]{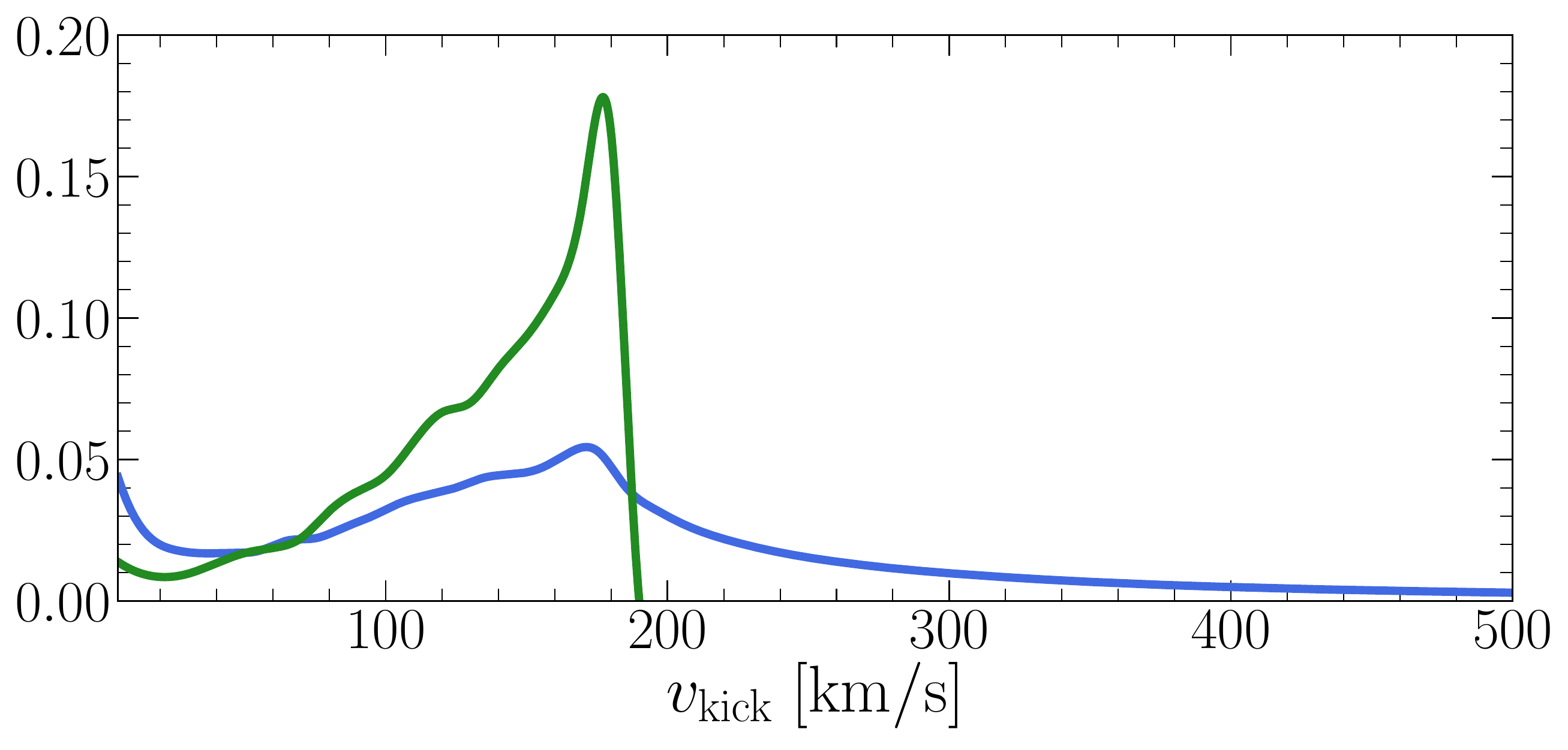}
  \caption{Kick imparted to the primary BH in GW190412, assuming that it originates from a previous BH merger. The blue curve represents the uniform-spin prior, while the green curve corresponds to the zero-spin prior. The median and 90\% credible interval are $v_{\rm kick}=158^{+240}_{-125}\;\rm km/s$ (uniform-spin prior) and $v_{\rm kick}=141^{+33}_{-71}\,\rm km/s$ (zero-spin prior).
  } 
\label{fig:vkhist}
\end{figure}

The results of this procedure are shown in Fig.~\ref{fig:mcorner}. We find that the parents of the primary BH in GW190412 
had masses $M_1=  25.24^{+3.55}_{-4.83} \Msun$ and $M_2=5.61^{+5.67}_{-3.98}\Msun$. The negative correlation between $M_1$ and $M_2$ apparent in Fig.~\ref{fig:mcorner} follows from the constraint $M_1+M_2\simeq m_1$, which is accurate up to small corrections due to the radiated energy $E_{\rm rad}\lesssim 5\%$.

Using the procedure that we just described, we can also estimate how much linear momentum was emitted during the merger of the parents, and hence the kick imparted 
to the primary component of GW190412 when it formed. Using the numerical-relativity fit assembled in Ref.~\cite{Gerosa:2016sys}, we find $v_{\rm kick}=158^{+240}_{-125}\;\rm km/s$.
 The resulting distribution
 is shown in Fig.~\ref{fig:vkhist}. The observed bimodality is a direct
 consequence of the distributions of $Q$ found in Fig.~\ref{fig:corner}. The peak at $Q\sim 0.2$ is responsible for the dominant mode at $v_{\rm kick}> 100$ km/s \cite{Gonzalez:2006md}, while the tail at $Q\gtrsim 0$ produces the secondary mode at $v_{\rm kick}\gtrsim 0$, as predicted in the test-particle limit.

This estimate implies that, if GW190412 is indeed a hierarchical merger, it must have formed in an environment with escape speed $v_{\rm esc}\gtrsim 100$ km/s (otherwise the primary BH would have been ejected, preventing the formation of GW190412 itself), in agreement with previous work by \citeauthor{Gerosa:2020bjb} \cite{Gerosa:2020bjb}. Astrophysical environments with such escape speeds include nuclear star clusters and AGN disks, but not globular clusters~\cite{Merritt:2004xa}, with their escape speeds ranging from $10$ to $100$ km/s \cite{Gnedin:2002un}. In particular, we find $p(v_{\rm kick}>10 {\rm km/s})=0.94$, $p(v_{\rm kick}>50 {\rm km/s})=0.87$, and $p(v_{\rm kick}>100 {\rm km/s})=0.75$.    %

Under a more restrictive---but arguably astrophysically motivated---prior where the progenitors are assumed to be nonspinning ($\xi_1=\xi_2=0$), we obtain a similar estimate for the mass ratio: $\qp=0.22_{-0.11}^{+0.14}$ (green line in the top panel of Fig.~\ref{fig:corner}). The masses and recoil velocities calculated for nonspinning progenitors are consistent with those obtained under uniform-spin priors, but with smaller errors: we find $M_1=25.24^{+2.35}_{-3.35}\Msun$, $M_2=5.44^{+3.41}_{-2.80}\Msun$ and $v_{\rm kick}=141^{+33}_{-71}\,\rm km/s$.

When interpreting our results, one should keep in mind that our error budget  takes into account only the statistical uncertainties on the parameters of the GW events. Much like anything else in science, our procedure also suffers from systematics (for instance, in the specific numerical-relativity fits used here).

\section{Conclusions}
\label{concl}

Most of the literature on compact binaries is concerned with determining the properties of a merger remnant from the properties of its progenitors. In this paper we addressed the inverse problem in two steps (see Fig.~\ref{cartoonparents}): 

\begin{enumerate}[label=(\alph*)]
\item  If we observe a binary system that ends up merging in LIGO, can we tell from the observed properties of the binary---and in particular from the effective-spin parameters $(\chieff,\,\chip)$----whether one of the BHs came from a previous (hierarchical) merger?
\item If indeed one of the BHs came from a previous merger, can we determine at least some of the properties of its parents?
\end{enumerate}

Given the parameters of a binary system, the direct problem has a well-defined solution that can be found by solving the field equations of general relativity. On the contrary, the inverse problem is in general overdetermined, because the only information available to us comes from the gravitational radiation emitted by the observed binary. We have demonstrated that this is enough to infer at least some of the properties of the parent binary. 

In particular, our key results are as follows:
\begin{enumerate}[label=(\alph*)]
\item Hierarchical mergers occupy a distinct region in the $(\chi_{\rm eff},\,\chi_{\rm p})$ plane. Therefore, a measurement of these spin parameters can add considerable weight to a hierarchical-merger interpretation of an event located in the pair-instability mass gap. Notably, this is the case for  GW190521.
\item If one of the BH binary components (not necessarily the primary) does indeed have hierarchical origin and its spin magnitude is well measured, one can place significant constraints on the properties of its parents. For example, we can infer that the parents of the heavier BH in GW190412 must have had a moderate mass ratio  $\qp=0.21_{-0.16}^{+0.32}$ and moderately low spins.
\end{enumerate}

We stress that, while we focused on the examples of GW190521 and GW190412, the formalism developed in this work is quite general.

The inversion procedure presented in Sec.~\ref{parents} requires estimates of the individual spin components. While challenging at the current detector sensitivity, these are bound to become routine in the future. For stellar-mass BH binaries,  third-generation ground-based detectors will measure component spin magnitudes within $\sim 1\%$ (see, e.g., \cite{Vitale:2016icu}). A similar accuracy is expected for supermassive BH binaries observed with the spaceborne interferometer LISA~\cite{Klein:2015hvg}. In that case, the BH parents problem is arguably even more relevant than in the LIGO context, given that supermassive BHs are firmly believed to grow hierarchically following the assembly of large-scale structures in the Universe \cite{Begelman:1980vb}.

Measurements of the masses, spins, and kicks of the progenitor of a hierarchical merger can be translated into constraints on their birthplace. In the LIGO context, for example, they can be used to infer bounds on the escape speeds and masses of the clusters that may have produced such mergers via dynamical interactions~\cite{Gerosa:2019zmo}. We have illustrated this point through a constraint of the cluster's escape speed for GW190412, but a more rigorous implementation of this idea requires better astrophysical modeling, which will be the subject of future work.

The identification of the parents' properties can be used to validate or reject the hierarchical origin of a GW event, because the inferred parents should presumably be part of the same population of merging BHs observed by our detectors. A posterior predictive check using the rest of the observed GW catalog could be used to verify this hypothesis (see e.g. \cite{Fishbach:2019ckx}). If {\it ad hoc} population outliers are required, this would rule out (or cast serious doubt on) the assumed hierarchical origin for the observed event. Such a test is a natural follow-up of this work.%

\acknowledgments

We thank Daria Gangardt for discussions.
V.B., E.B. and K.W.K.W. are supported by NSF Grants No. PHY-1912550 and AST-2006538, NASA ATP Grants No. 17-ATP17-0225 and 19-ATP19-0051, NSF-XSEDE Grant No. PHY-090003, and NSF Grant PHY-20043.
D.G. and M.M. are supported by European Union's H2020  ERC Starting Grant No. 945155--GWmining  and Royal Society Grant No. RGS-R2-202004.
D.G. is supported by Leverhulme Trust Grant No. RPG-2019-350. 
We would like to acknowledge networking support by the GWverse COST Action CA16104, ``Black holes, gravitational waves and fundamental physics.'' Computational work was performed on the University of Birmingham BlueBEAR cluster and the Maryland Advanced Research Computing Center (MARCC).

\vspace{0cm}
\bibliography{remnant}

\end{document}